\documentclass[10pt,conference]{IEEEtran}
\usepackage{amsmath}
\usepackage{subfigure}
\usepackage{epsfig}
\usepackage{amssymb}
\usepackage{float}
\usepackage{setspace}
\usepackage{hyperref}
\usepackage{tabularx,booktabs}
\usepackage{url}
\usepackage{comment}
\usepackage[inline]{enumitem}
\usepackage{siunitx}

\usepackage[utf8]{inputenc}
\newcolumntype{H}{>{\setbox0=\hbox\bgroup}c<{\egroup}@{}}
\newcolumntype{G}{>{\collectcell\@gobble}c<{\endcollectcell}@{}}

\title{GitHub OSS Governance File Dataset}
\author{
\IEEEauthorblockN{Yibo Yan}
\IEEEauthorblockN{UC Davis\\
ybyan@ucdavis.edu}
\and
\IEEEauthorblockN{Seth Frey}
\IEEEauthorblockN{UC Davis\\
sethfrey@ucdavis.edu}
\and
\IEEEauthorblockN{Amy Zhang}
\IEEEauthorblockN{UW Seattle\\
axz@cs.uw.edu}
\and
\IEEEauthorblockN{Vladimir Filkov\\}
\IEEEauthorblockN{UC Davis\\
vfilkov@ucdavis.edu}
\and
\IEEEauthorblockN{Likang Yin}
\IEEEauthorblockN{UC Davis\\
lkyin@ucdavis.edu}
}

\begin{document}
\maketitle
\begin{abstract}
Open-source Software (OSS) has become a valuable resource in both industry and academia over the last few decades.
Despite the innovative structures they develop to support the projects, OSS projects and their communities have complex needs and face risks such as getting abandoned. 
To manage the internal social dynamics and community evolution, 
OSS developer communities have started relying on written governance documents that assign roles and responsibilities to different community actors.

To facilitate the study of the impact and effectiveness of formal governance documents on OSS projects and communities, we present a longitudinal dataset of 710 GitHub-hosted OSS projects with \path{GOVERNANCE.MD} governance files. This dataset includes all commits made to the repository, all issues and comments created on GitHub, and all revisions made to the governance file. We hope its availability will foster more research interest in studying how OSS communities govern their projects and the impact of governance files on communities.
\end{abstract}

\section{Introduction}
\label{intro}
As one of the largest online open code hosting platforms, GitHub provides a widely accessible and easy-to-use platform for many Open Source Software (OSS) projects and communities. With the free availability of OSS digital traces, studying OSS at scale has become feasible, and academic studies of OSS repositories have proliferated. 

Research into OSS communities spans topics including code quality~\cite{lu2016does,stuanciulescu2022code}, sustainability~\cite{gamalielsson2014sustainability,yin2021sustain,chengalur2010sustainability}, and more recently governance~\cite{izquierdo2015enabling,li2021code}.
GitHub governance is different than the governance in other OSS hosting organizations. For instance, Apache Software Foundation (ASF), a well-known not-for-profit organization, provides an incubator for OSS projects. Projects in the ASF Incubator (ASFI) receive mentorship, and their efforts are guided and governed by ASFI's committee. Most OSS projects and communities on GitHub do not have such a committee to support OSS activities. As a solution, many OSS communities draft their own governance files and utilize GitHub's ``issue'' feature to govern and coordinate OSS activities. Based on the existing convention of using text-based documentation to govern collective activities (e.g., Contributor Covenant), open-source communities start to adopt the approach of hosting a text-based file named \path{GOVERNANCE.MD}, making an effort to foster a healthier community and convey labor division and participation expectations more clearly.
As a project evolves, the governance file often evolves through periodic revisions, as community members introduce, modify, or delete governance rules.

Studying governance files can shed light on how OSS projects coordinate work and potentially reveal  aspects of the socio-technical dynamics within OSS projects. Understanding how OSS projects on GitHub govern their collaborative work and how socio-technical dynamics change within projects can further facilitate the study of sustainability in OSS projects to pinpoint crucial factors in the successful governance of OSS projects and communities \cite{amrollahi2014open,ghapanchi2011taxonomy}. 

We present the GitHub Open-Source Software governance documentation dataset. It includes governance files, projects' commit history, and issues for 710 OSS projects hosted on GitHub. To facilitate  longitudinal studies, we provide separate tables, capturing the commit history of \path{GOVERNANCE.MD} files and detailed changes made in each commit on the \path{GOVERNANCE.MD} file at line-level granularity. 

To the best of our knowledge, this is the first time such a governance-documentation-oriented GitHub-hosted OSS project dataset has been presented to the empirical software engineering community.
Next, we present the details of our dataset, scraping methodology, and storage, followed by two preliminary examples of research studies that can benefit from this data. Our dataset, along with the scripts we used to scrape it, is available at Zenodo: \href{https://doi.org/10.5281/zenodo.7530768}{https://doi.org/10.5281/zenodo.7530768}.

\section{Related Work}
\label{sec:priorwork}
Analyzing the community organization, structure, and governance of OSS projects from a socio-technical perspective enables the understanding of dynamics~\cite{ducheneaut2005socialization}, project quality over time~\cite{bird2009putting}, and project coordination effectiveness~\cite{bird2011sociotechnical} within OSS projects~\cite{herbsleb2007global,scacchi2005socio}.
Besides GitHub, many large organizations and foundations like OSGeo \cite{osgeo_2022}, Apache Software Foundation \cite{yin2022open,yin2021apache}, and Linux Foundation \cite{linux_2022} also support a multitude of OSS projects. Various tools have been produced to mine these projects, e.g., Perceval provides a unified entry-point to gather the data of software repositories from various backends~\cite{duenas2018perceval}; GHTorrent provides a streamlined approach to gather mirrored data from GitHub~\cite{Gousi13}; Project CHAOSS delivers metrics, model and software to  understand OSS community health~\cite{chaoss_2022}. 
Our governance documentation dataset is orthogonal, and together with other sources can enable holistic longitudinal analyses of OSS project evolution.

\section{OSS Governance Documentation Dataset}

The dataset comprises a longitudinal record of the entire revision history of the \path{GOVERNANCE.MD} file for each collected project, at varying levels of granularity. Besides the detailed information on each commit that developers made to the governance file, we also extracted the line change information for each commit made on the governance file. 

\subsubsection{Data Access}
GitHub maintains a set of interfaces through which developers can fetch repository-related data. GitHub supports both REST \cite{githubrest_2022} and GraphQL \cite{githubgraphql_2022} APIs for data retrieval, and we take advantage of both to scrape the data. Since GitHub provides first-hand data access, the data gathered through GitHub's official APIs is at least as reliable and up-to-date as the data from other social coding sites. 
Although GHTorrent serves the purpose of mining GitHub, it might not be useful for particular research directions, as it produces excessive data by exhaustively mirroring GitHub data but misses detailed data on the governance file's per-line revisions. 


\subsubsection{Repositories}
Based on GitHub search results, there are more than 1.6M projects with a \path{GOVERNANCE.MD} file. However, not all of them are signaling meaningful information as these projects often do not have a \path{GOVERNANCE.MD} file for their own projects but include other packages or dependencies that have \path{GOVERNANCE.MD} files. 
~\footnote{This is because the npm installs packages/dependencies in the project directory, and project developers do not ``gitignore" the dependency folder.} 
Therefore, to reduce noise in the data set,
we used the REST API's search code endpoint~\footnote{\url{https://docs.github.com/en/rest/search?apiVersion=2022-11-28\#search-code}} 
to fetch only the repositories that contain a file with a \path{GOVERNANCE.MD} (case-insensitive and non-exact match) in the root directory.~\footnote{Therefore, results like `Governance-Committee\_Charter.md', `IPEP-29:-Project-Governance.md', and `GovernancePolicy.md', etc., exist.}
The initial data set consists of 1,899 unique projects that have a \path{GOVERNANCE.MD} file in their root directory.
To ensure the projects are meaningful, we only keep projects that have at least 1 commit/issue.
This results in the final data set of 710 projects.
We gathered a list of basic metadata for each repository, such as the repository name and the filename of the governance file.


Due to the non-deterministic nature of GitHub's search API, we kept calling the search API repetitively until 99 out of the 100 last searched repositories were already encountered. 
We used {\textit{gql}}\footnote{\url{https://github.com/graphql-python/gql}} to fetch commits, issues, and comments in each issue for each repository. Additionally, we used \textit{PyDriller} \cite{pydriller2018} to locally locate all commits that involve a change on the governance file. With the help of \textit{PyDriller}, we obtained detailed information on each commit made on the governance file at the granularity of a line by processing through the parsed diff output, offering a view of which line along with the edited content is added or deleted in each commit. 

\subsubsection{Commits, Issues, and Comments}
We used the GraphQL endpoint, \url{https://docs.github.com/en/graphql/guides/forming-calls-with-graphql\#the-graphql-endpoint}, to collect commits, issues, and comments inside each issue.
We sent sequential GraphQL queries to iterate through the search space by adjusting the cursor correspondingly.

Preliminary project-scoped aggregations were done on the collected data to generate basic metrics for each repository and were added to the repository list afterward. As a result, the repository list contains information regarding the number of stars, forks, commits, committers, issues, and comments; the number of submitters of issues and comments; and the number of commits and committers on the governance file.

\subsubsection{Commit History of the Governance File}
As GitHub lacks support for retrieving commit history of a specific file,
we used PyDriller \cite{pydriller2018} to examine commits that involve a change (i.e., add, delete, or modify) to the governance file. We extracted basic metadata, e.g., \textit{author}, \textit{committer}, \textit{fileLOC}, along with the content and diff output before and after the commit. We also extracted information on how each line got modified (i.e., added or deleted) in each commit on the governance file.
We used scripts on the collected list of commits on the governance file and obtained: a) a list of commits per section of the governance file;
\footnote{The section is defined as the content between two heading elements in the Markdown file. The heading element is written as `\#', `\#\#', etc., followed by the heading content.}
and b) a list of the latest governance file of every repository by pulling out the content after the last collected commit on the governance file.
All data---commits, issues, comments---were collected from the initial creation date of the corresponding project on GitHub to June 2022, the last day updated by the script.

\begin{figure*}[!htbp]
  \centering
   \includegraphics[width=7.1in]{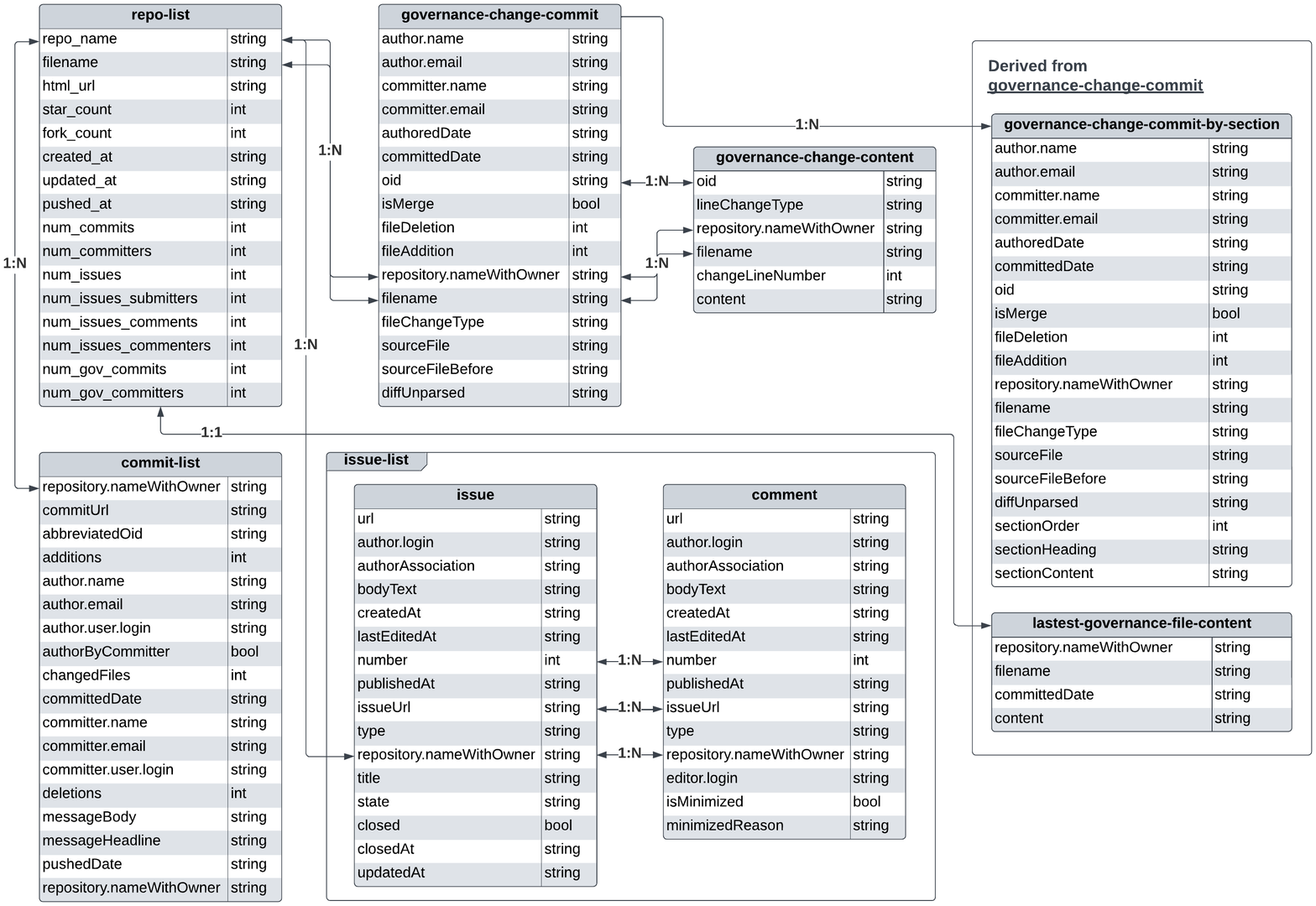}
\caption{Data Schema of the GitHub OSS Governance Documentation Dataset. Note: the actual data type can vary based on different sources for the archives, e.g., for a MongoDB dump, all date fields like `createdAt' will be a native \textit{Date} type in the MongoDB context.}
\label{schema}
\end{figure*}

\subsubsection{Database}
\label{sec:storage}
Most data were stored directly in a MongoDB database, a type of NoSQL database, and later were exported to different formats of data, i.e., CSV, SQL, and MongoDB archive dumps. A small portion of data, such as the repository list, was initially processed by \textit{pandas} in-memory and written to the MongoDB database afterward by scripts. All data were re-exported in the end. We used {PyMongo}\footnote{\url{https://github.com/mongodb/mongo-python-driver}} to interact with the MongoDB database via python scripts. 
The associated data schema is shown in \autoref{schema}.
Each table in the schema corresponds to one CSV table, SQL table, or MongoDB collection in the dataset.

Table \textit{\textbf{repo-list}} contains a list of metadata of GitHub-hosted projects. Each row represents a project. There is no explicit foreign key across tables; instead, \textit{`repo\_name'} and \textit{`filename'} serve as two major keys to locate commits, issues, and comments of the corresponding repository. 


Table \textit{\textbf{commit-list}} contains the commits from all projects in  \textbf{\textit{repo-list}}; each row represents a commit.

Table \textit{\textbf{governance-change-commit}} contains the commits for the governance file of every project. Each row represents a commit. As this list was collected by locally analyzing git repositories, no GitHub-related information is presented. 


Table \textit{\textbf{governance-change-content}} contains the list of line-change information of each commit made on the governance file over all collected projects. Each row represents one line-change information of a specific commit. The combination of \textit{oid}, \textit{repository.nameWithOwner}, and \textit{filename} should be used as the key to locate all line changes on one specific commit. 


Table \textit{\textbf{governance-change-commit-by-section}} contains the same information as \textbf{\textit{governance-change-commit}}. For each row in the \textbf{\textit{governance-change-commit}} table, the \textit{`sourceFile'} field was used to extract section information. Each section information was stored as a row in the \textit{\textbf{governance-change-commit-by-section}} table. Most fields are identical to the fields in the \textbf{\textit{governance-change-commit}} table. 


Table \textit{\textbf{lastest-governance-file-content}} contains the latest version of the governance file of each collected project. 


\subsubsection{Metrics}
We compute a set of standard software engineering metrics for OSS project activity from our governance documentation dataset. These include: 
the number of stars (\textit{star\_count}), 
forks (\textit{fork\_count}),
commits (\textit{num\_commits}),
committers (\textit{num\_committers}),
issues (\textit{num\_issues}),
issue submitters (\textit{num\_issues\_submitters}),
and issue comments (\textit{num\_issues\_comments}); 
the number of people who commented on issues (\textit{num\_issues\_commenters});
the number of commits to the governance file (\textit{num\_gov\_commits}); and
the number of people committed to it (\textit{num\_gov\_committers}).
The descriptive statistics for these metrics over the 710 GitHub-hosted projects are given in Table~\ref{statitics}.

\begin{table}[t] \centering 
  \caption{The project-level statistics of the 710 GitHub-hosted projects with the governance markdown file}
  \label{statitics}
    \scalebox{0.8}{
        \begin{tabular}{lccccc}
        \\[-1.8ex]\hline 
        \hline \\[-1.8ex] 
        Statistic & \multicolumn{1}{c}{Mean} &  \multicolumn{1}{c}{Median} & \multicolumn{1}{c}{St. Dev.} & \multicolumn{1}{c}{Min} & \multicolumn{1}{c}{Max} \\ 
        \hline \\[-1.8ex] 
        star\_count             & 1500.8   & 42.5   & 5902.7   & 0    & 85775 \\ 
        fork\_count             & 308.3    & 23     & 1305.3   & 0    & 22861 \\ 
        num\_commits            & 2,865.3  & 350.5  & 6,896.2  & 0    & 60,612 \\ 
        num\_committers         & 164.5    & 19     & 495.7    & 0    & 4,033 \\ 
        num\_issues             & 419.2    & 43     & 1,453.8  & 1    & 23,609 \\ 
        num\_issues\_submitters & 149.34   & 12     & 629.8    & 1    & 10,069 \\ 
        num\_issues\_comments   & 1,344    & 71     & 5,724.3  & 0    & 83,183 \\ 
        num\_issues\_commenters & 168.9    & 13     & 842.1    & 0    & 14,285 \\ 
        num\_gov\_commits       & 5.5      & 2      & 9.1      & 1    & 79 \\ 
        num\_gov\_committers    & 2.4      & 1      & 2.8      & 1    & 26 \\ 
        \hline \\[-1.8ex] 
        \end{tabular}
    }
\end{table} 

\section{Potential Use Cases}
This section presents two potential use cases that can be studied based on the presented dataset.

\subsection{Case I: Studying Governance in Digital Commons}
Scholars of self-organized self-governance in the online context have long drawn on the theories of common pool resource scholars such as Elinor Ostrom~\cite{schweik2012internet,frey2019emergence}. One particularly prominent contribution from this literature are the Design Principles for Sustainable Common Pool Resource Management~\cite{ostrom2015governing}. These were developed for studying  institutions for managing natural resources such as fisheries, forests, and water systems, but have been extended to OSS and other digital resources~\cite{frey2019place}. 
Yet, there remains a critical need to test the generality of the design principles in the OSS context, which would require a comprehensive comparative dataset of formal OSS governance records. 
For example, with a large collection of formal governance documents on GitHub, a scholar could test ``Principle 1: Clearly defined boundaries" by extracting reference to GitHub platform position constructs like users, committers, and contributors. For ``Principle 5: Graduated sanctions for appropriators who do not respect community rules" a scholar might search \path{GOVERNANCE.md} files for whether the rules contain a variety of sanctions (both warnings and bans, not just one or the other).
With this dataset, we can apply these principles in any open-source project hosted on GitHub and see how it works and whether the best practices align with these principles. 
We can also study the collaboration mechanisms, the rules and norms set in the community, and the role of institutions, such as maintainers, collaborators, and contributors in the governance of the project.

\subsection{Case II: Semantic Search in Institutions Analysis}
Institutional analysis using natural language processing (NLP) methods on GitHub can be used to study the organizational structure, rules, and norms of online communities, and how they relate to the governance and performance of open-source software projects.
With the classified sentences, we can conduct institutional analysis to understand the governance mechanisms. In what way do people invoke the rules of an open-source project in regular discourse?  What is the overlap between the rules people discuss and those they use?  How do rules change over time, and in what direction? How many people tend to contribute to rules, and what is the typical number of discrete roles they define as a project matures?  These questions all inform the practice of OSS governance, and effective peer production generally.  The dataset we introduce thus makes it possible to advance the entire front of online governance research.

\subsection{Case III: Development of Governance Design Patterns}
By examining a repository of enacted governance across a range of communities, governance practitioners and tool developers can gain an understanding of what governance components are common. Broadly common components could be surfaced as governance design patterns to then be integrated into guides and other materials to help communities who are deciding on what kind of governance to enact~\cite{schneider2020designing}. These patterns could also be translated into programs within existing software toolkits for programmatically enacting governance~\cite{zhang2020policykit}.
Finally, developers of governance-related features on GitHub or third-party tools could build in customized support for common governance components; for instance, implementing certain roles, tiered permissions, or voting mechanisms that were often expressed.

\section{Limitations and Conclusion}\label{sec:limit}
\underline{Limitations} We note that this dataset has two major limitations regarding \textbf{completeness} and potential \textbf{bias}.
    \textbf{Completeness} 
        \begin{enumerate*}
            \item Because GitHub's search API doesn't return a deterministic and full set of results, the dataset is not a complete set of all GitHub-hosted repositories with a governance file.
            \item As we only collected projects which contain the \path{GOVERNANCE.MD} file in the root directory, some GitHub-hosted projects are missing from our dataset as they might organize and store their governance files differently. For example, some projects put their governance files directly in the \textit{readme} file.
            \item The commit history collected in our dataset does not necessarily represent the full history of all commits that have been made. Certain Git operations may lose commit history, such as \textit{squash} or \textit{rebase}.
        \end{enumerate*}
    \textbf{Bias}
        \begin{enumerate*}
            \item Some projects may use the same governance file. Under some circumstances, multiple projects may put the same line of redirection link in their project-level governance file, referring to the external governance documentation. In this case, all governance files in these projects will produce the exact same but not meaningful governance content.
            \item Some projects didn't start at GitHub in the first place; instead, the projects and corresponding governance files were hosted on other platforms and were moved to GitHub in the later development stage. In this case, the commit history on the governance file might be incomplete, causing bias in analytic studies.
        \end{enumerate*}

To mitigate the completeness and bias mentioned above, we can 1) integrate data from other sources, such as GHTorrent's mirrored data, to workaround GitHub APIs' non-deterministic behavior; 
2) we can use the available governance file dataset to train a classifier to classify whether some text files contain governance content, avoiding searching for the governance file in a specific pattern and expanding the search space to include more repositories.

\underline{Conclusion} 
This work presents the development of a longitudinal dataset of 710 Open Source Software (OSS) projects hosted on GitHub that includes information about governance files. OSS projects and communities still fail frequently, despite the popularity of hosting platforms like GitHub, and governance files are sometimes drafted and revised to serve the community's needs.

This dataset aims to help researchers and developers understand best practices and common patterns in OSS governance documentation and to identify projects with poor governance that could lead to better maintainability and sustainability in the long run. 
We present how the data was collected and what specific criteria were used to select the projects, as well as a description of the data, its structure, and the challenges encountered during the data collection process. Additionally, we discussed the potential applications and the value of the dataset and encourage researchers to use this dataset to study the state of open-source software governance across different projects and communities.


\section*{Acknowledgement}
We thank the anonymous reviewers for their constructive
comments. This material is based upon work supported by the
National Science Foundation under GCR \#2020751/2020900 ``Jumpstarting Successful OSS Projects With Evidence-Based Rules and Structures", and DASS \#2217652/2217653 ``Transitioning OSS projects to accountable community governance".

\bibliographystyle{IEEEtra}
\bibliography{ref}

\begin{thebibliography}{10}
\providecommand{\url}[1]{#1}
\csname url@samestyle\endcsname
\providecommand{\newblock}{\relax}
\providecommand{\bibinfo}[2]{#2}
\providecommand{\BIBentrySTDinterwordspacing}{\spaceskip=0pt\relax}
\providecommand{\BIBentryALTinterwordstretchfactor}{4}
\providecommand{\BIBentryALTinterwordspacing}{\spaceskip=\fontdimen2\font plus
\BIBentryALTinterwordstretchfactor\fontdimen3\font minus
  \fontdimen4\font\relax}
\providecommand{\BIBforeignlanguage}[2]{{%
\expandafter\ifx\csname l@#1\endcsname\relax
\typeout{** WARNING: IEEEtran.bst: No hyphenation pattern has been}%
\typeout{** loaded for the language `#1'. Using the pattern for}%
\typeout{** the default language instead.}%
\else
\language=\csname l@#1\endcsname
\fi
#2}}
\providecommand{\BIBdecl}{\relax}
\BIBdecl

\bibitem{lu2016does}
Y.~Lu, X.~Mao, Z.~Li, Y.~Zhang, T.~Wang, and G.~Yin, ``Does the role matter? an
  investigation of the code quality of casual contributors in github,'' in
  \emph{2016 23rd Asia-Pacific Software Engineering Conference (APSEC)}.\hskip
  1em plus 0.5em minus 0.4em\relax IEEE, 2016, pp. 49--56.

\bibitem{stuanciulescu2022code}
S.~St{\u{a}}nciulescu, L.~Yin, and V.~Filkov, ``Code, quality, and process
  metrics in graduated and retired asfi projects,'' in \emph{Proceedings of the
  30th ACM Joint European Software Engineering Conference and Symposium on the
  Foundations of Software Engineering}, 2022, pp. 495--506.

\bibitem{gamalielsson2014sustainability}
J.~Gamalielsson and B.~Lundell, ``Sustainability of open source software
  communities beyond a fork: How and why has the libreoffice project evolved?''
  \emph{Journal of Systems and Software}, vol.~89, pp. 128--145, 2014.

\bibitem{yin2021sustain}
\BIBentryALTinterwordspacing
L.~Yin, Z.~Chen, Q.~Xuan, and V.~Filkov, ``Sustainability forecasting for
  apache incubator projects,'' ser. ESEC/FSE 2021.\hskip 1em plus 0.5em minus
  0.4em\relax New York, NY, USA: Association for Computing Machinery, 2021, p.
  1056–1067. [Online]. Available:
  \url{https://doi.org/10.1145/3468264.3468563}
\BIBentrySTDinterwordspacing

\bibitem{chengalur2010sustainability}
I.~Chengalur-Smith, A.~Sidorova, and S.~L. Daniel, ``Sustainability of
  free/libre open source projects: A longitudinal study,'' \emph{Journal of the
  Association for Information Systems}, vol.~11, no.~11, p.~5, 2010.

\bibitem{izquierdo2015enabling}
J.~L.~C. Izquierdo and J.~Cabot, ``Enabling the definition and enforcement of
  governance rules in open source systems,'' in \emph{2015 IEEE/ACM 37th IEEE
  International Conference on Software Engineering}, vol.~2.\hskip 1em plus
  0.5em minus 0.4em\relax IEEE, 2015, pp. 505--514.

\bibitem{li2021code}
R.~Li, P.~Pandurangan, H.~Frluckaj, and L.~Dabbish, ``Code of conduct
  conversations in open source software projects on github,'' \emph{Proceedings
  of the ACM on Human-computer Interaction}, vol.~5, no. CSCW1, pp. 1--31,
  2021.

\bibitem{amrollahi2014open}
A.~Amrollahi, M.~Khansari, and A.~Manian, ``How open source software succeeds?
  a review of research on success of open source software,'' 2014.

\bibitem{ghapanchi2011taxonomy}
A.~H. Ghapanchi, A.~Aurum, and G.~Low, ``A taxonomy for measuring the success
  of open source software projects,'' \emph{First Monday}, vol.~16, no.~8,
  2011.

\bibitem{ducheneaut2005socialization}
N.~Ducheneaut, ``Socialization in an open source software community: A
  socio-technical analysis,'' \emph{Computer Supported Cooperative Work
  (CSCW)}, vol.~14, no.~4, pp. 323--368, 2005.

\bibitem{bird2009putting}
C.~Bird, N.~Nagappan, H.~Gall, B.~Murphy, and P.~Devanbu, ``Putting it all
  together: Using socio-technical networks to predict failures,'' in \emph{2009
  20th International Symposium on Software Reliability Engineering}.\hskip 1em
  plus 0.5em minus 0.4em\relax IEEE, 2009, pp. 109--119.

\bibitem{bird2011sociotechnical}
C.~Bird, ``Sociotechnical coordination and collaboration in open source
  software,'' in \emph{2011 27th IEEE International Conference on Software
  Maintenance (ICSM)}.\hskip 1em plus 0.5em minus 0.4em\relax IEEE, 2011, pp.
  568--573.

\bibitem{herbsleb2007global}
J.~D. Herbsleb, ``Global software engineering: The future of socio-technical
  coordination,'' in \emph{Future of Software Engineering (FOSE'07)}.\hskip 1em
  plus 0.5em minus 0.4em\relax IEEE, 2007, pp. 188--198.

\bibitem{scacchi2005socio}
W.~Scacchi, ``Socio-technical interaction networks in free/open source software
  development processes,'' in \emph{Software process modeling}.\hskip 1em plus
  0.5em minus 0.4em\relax Springer, 2005, pp. 1--27.

\bibitem{osgeo_2022}
\BIBentryALTinterwordspacing
Apr 2022. [Online]. Available: \url{https://www.osgeo.org/}
\BIBentrySTDinterwordspacing

\bibitem{yin2022open}
L.~Yin, M.~Chakraborti, Y.~Yan, C.~Schweik, S.~Frey, and V.~Filkov, ``Open
  source software sustainability: Combining institutional analysis and
  socio-technical networks,'' \emph{Proceedings of the ACM on Human-Computer
  Interaction}, vol.~6, no. CSCW2, pp. 1--23, 2022.

\bibitem{yin2021apache}
L.~Yin, Z.~Zhang, Q.~Xuan, and V.~Filkov, ``Apache software foundation
  incubator project sustainability dataset,'' in \emph{2021 IEEE/ACM 18th
  International Conference on Mining Software Repositories (MSR)}.\hskip 1em
  plus 0.5em minus 0.4em\relax IEEE, 2021, pp. 595--599.

\bibitem{linux_2022}
\BIBentryALTinterwordspacing
 [Online]. Available: \url{https://www.linuxfoundation.org/}
\BIBentrySTDinterwordspacing

\bibitem{duenas2018perceval}
S.~Due{\~n}as, V.~Cosentino, G.~Robles, and J.~M. Gonzalez-Barahona,
  ``Perceval: software project data at your will,'' in \emph{Proceedings of the
  40th International Conference on Software Engineering: Companion
  Proceeedings}.\hskip 1em plus 0.5em minus 0.4em\relax ACM, 2018, pp. 1--4.

\bibitem{Gousi13}
\BIBentryALTinterwordspacing
G.~Gousios, ``The ghtorrent dataset and tool suite,'' in \emph{Proceedings of
  the 10th Working Conference on Mining Software Repositories}, ser. MSR
  '13.\hskip 1em plus 0.5em minus 0.4em\relax Piscataway, NJ, USA: IEEE Press,
  2013, pp. 233--236. [Online]. Available:
  \url{http://dl.acm.org/citation.cfm?id=2487085.2487132}
\BIBentrySTDinterwordspacing

\bibitem{chaoss_2022}
\BIBentryALTinterwordspacing
 [Online]. Available: \url{https://chaoss.community/}
\BIBentrySTDinterwordspacing

\bibitem{githubrest_2022}
\BIBentryALTinterwordspacing
 [Online]. Available:
  \url{https://docs.github.com/en/rest?apiVersion=2022-11-28}
\BIBentrySTDinterwordspacing

\bibitem{githubgraphql_2022}
\BIBentryALTinterwordspacing
 [Online]. Available: \url{https://docs.github.com/en/graphql}
\BIBentrySTDinterwordspacing

\bibitem{pydriller2018}
D.~Spadini, M.~Aniche, and A.~Bacchelli, \emph{PyDriller: Python Framework for
  Mining Software Repositories}, 2018.

\bibitem{schweik2012internet}
C.~M. Schweik and R.~C. English, \emph{Internet success: a study of open-source
  software commons}.\hskip 1em plus 0.5em minus 0.4em\relax MIT Press, 2012.

\bibitem{frey2019emergence}
S.~Frey and R.~W. Sumner, ``Emergence of integrated institutions in a large
  population of self-governing communities,'' \emph{PloS one}, vol.~14, no.~7,
  p. e0216335, 2019.

\bibitem{ostrom2015governing}
E.~Ostrom, \emph{Governing the Commons: The Evolution of Institutions for
  Collective Action}, ser. Canto Classics.\hskip 1em plus 0.5em minus
  0.4em\relax Cambridge University Press, 2015.

\bibitem{frey2019place}
S.~Frey, P.~Krafft, and B.~C. Keegan, ``" this place does what it was built
  for" designing digital institutions for participatory change,''
  \emph{Proceedings of the ACM on Human-Computer Interaction}, vol.~3, no.
  CSCW, pp. 1--31, 2019.

\bibitem{schneider2020designing}
N.~Schneider, ``Designing community self-governance with communityrule,'' 2020.

\bibitem{zhang2020policykit}
A.~X. Zhang, G.~Hugh, and M.~S. Bernstein, ``Policykit: building governance in
  online communities,'' in \emph{Proceedings of the 33rd Annual ACM Symposium
  on User Interface Software and Technology}, 2020, pp. 365--378.

\end{thebibliography}

\end{document}